\documentclass[10pt,fleqn]{article}
\usepackage{amsmath,graphicx,latexsym}

\begin{document}

\title{Three Dimensional Integrable Mappings}

\author{Apostolos Iatrou \thanks{{\copyright\mbox{ }Apostolos Iatrou}}
\thanks{Part of this paper is taken from the author's PhD thesis (La Trobe University).
The part taken was written under the supervision of K. A. Seaton.}
\thanks{{\sf email:
A.Iatrou@latrobe.edu.au\mbox{ }\mbox{ }apostolosiatrou@hotmail.com
}}}

\date{}

\maketitle

\begin{abstract}

We derive three-dimensional integrable mappings which have two
invariants.

\end{abstract}

\section{Introduction}

In this paper we focus on three-dimensional integrable autonomous
mappings preserving at least one three-quadratic (possibly
rational) integral (we considered the four-dimensional case
(together with its generalization to higher dimensions) in
\cite{iat}). A major reason for such a study is the lack off
results on three-dimensional integrable mappings. A recent paper,
which makes some progress on three-dimensional integrable
mappings, is \cite{hky}. In this paper Hirota {\em et al} used
algebraic entropy \cite{hv} to determine which three-dimensional
mappings of a particular form had polynomial growth implying zero
algebraic entropy. Having discovered all such possible mappings,
they used a procedure outlined in their paper to find two
functionally independent conserved quantities for each map. In
this paper we will take a different approach to the one used by
Hirota {\em et al} to construct three-dimensional integrable
mappings. For the purposes of this paper, we consider a
three-dimensional autonomous mapping integrable if there exist two
functionally independent integrals in involution with respect to
some Poisson structure.

The plan of this paper is as follows : In Section 2 we derive
three-dimensional volume-preserving mappings which preserve a
three-quadratic expression (a method introduced in \cite{casa} on
a rational four-quadratic expression), then assuming that these
three-dimensional volume-preserving mappings have a second
integral with a particular ansatz we find 3 three-dimensional
volume-preserving integrable mappings. In Section 3 we use the
processes of {\em reparametrization} and {\em replacement}
\cite{aijr1,aijr2,riq} (terms introduced and defined in
\cite{riq}) to construct three-dimensional measure-preserving
integrable mappings.

\section{Three-Dimensional Volume-Preserving Mappings}

In this section we construct three-dimensional volume-preserving
mappings (orientation-reversing and -preserving)\footnote{A
mappings $L$ is orientation reversing (orientation preserving) if
$\det dL =-1$ ($\det dL =1$).} possessing two integrals, at least
one of the integrals being quadratic in the three variables.

We begin with the orientation-reversing case. Consider the
three-quadratic expression
\begin{equation}\label{threequad}
  I(x,y,z)=\sum A_{\alpha_1 \beta_1 \gamma_1}
    \, \,x^{\alpha_1} y^{\beta_1} z^{\gamma_1}, \, \; \;
    (\alpha_1,\beta_1,\gamma_1=0,1,2),
\end{equation}
where $A_{\alpha_1\beta_1\gamma_1}$ are independent parameters.
Assume that (\ref{threequad}) is invariant under a cyclic
permutation of variables\footnote{This guarantees that the mapping
preserving this integral takes the form $x'=y,y'=z,z'=F(x,y,z)$,
where $F$ is some function.}, i.e. $I(x,y,z)=I(y,z,x)$, and that
the mapping, $L$, preserving $I(x,y,z)$ is reversible, i.e.
\begin{equation}\label{rev}
  L \circ G \circ L = G,
\end{equation}
with reversing symmetry $G : x'=y,y'=x,z'=x$. The reversing
symmetry also implies that $I(x,y,z)=I(z,y,x)$. Under these
conditions we obtain the integral
\begin{eqnarray}\label{cycinvint}
  I(x,y,z)&=&A_1 x^2 y^2 z^2+A_2
  xyz(xy+xy+yz)+A_3(x^2y^2+x^2z^2+y^2z^2)\nonumber \\
  & &{}+A_4 xyz(x+y+z)
  +A_5(x^2y+x^2z+xy^2+xz^2+y^2z+yz^2)\nonumber \\
  & &{}+A_6 xyz+A_7(x^2+y^2+z^2)+A_8(xy+xz+yz)+A_9(x+y+z).
\end{eqnarray}
The mapping, $L$, which leaves the integral (\ref{cycinvint})
invariant can be derived by setting $x'=x$ and $y'=y$ and
differencing the integral (\ref{cycinvint}), i.e.
$I(x',y',z')=I(x,y,z')=I(x,y,z)$. Then assuming $z'\ne z$, we can
solve for $z'$, to obtain the involution $L_z$. Finally, composing
$L_z$ with the cyclic shift $L_c:x'=y,y'=z,z'=x$, i.e. $L=L_z
\circ L_c$, we obtain the non-trivial volume-preserving
orientation-reversing mapping, $L$,
\begin{eqnarray}\label{mapt}
  x'&=&y \nonumber \\
  y'&=&z \nonumber \\
  z'&=&-x-\frac{A_2 y^2z^2+A_4 yz(y+z)+A_5(y^2+z^2)+A_6
  yz+A_8(y+z)+A_9}
  {A_1 y^2z^2+A_2 yz(y+z)+A_3(y^2+z^2)+A_4
  yz+A_5(y+z)+A_7}.
\end{eqnarray}
For a slightly more general map see \cite{gm}.

Next, assume that the mapping (\ref{mapt}) has a second integral
with the following ansatz\footnote{See the Appendix for the
reason why the ansatz has this form.}
\begin{equation}\label{ansatz}
  I(x,y,z)=\sum A_{\alpha_2 \beta_2 \gamma_2}
    \, \,x^{\alpha_2} y^{\beta_2} z^{\gamma_2}, \, \; \;
    (\alpha_2,\gamma_2=0,1,2\, ,\; \; \beta_2=0,1,2,3,4).
\end{equation}
where $A_{\alpha_2\beta_2\gamma_2}$ are independent parameters. As
the mapping (\ref{mapt}) is reversible we also have
$I_2(x,y,z)=I_2(z,y,x)$. We have found the following mappings
which simultaneously preserve integrals of the form
(\ref{cycinvint}) and (\ref{ansatz}) :
\begin{eqnarray}\label{map1}
  x'&=&y \nonumber \\
  y'&=&z \\
  z'&=&-x-
  \frac{\beta(y+z)^2+\epsilon(y+z)+\xi}
  {\beta(y+z)+\gamma}\nonumber
\end{eqnarray}
with integrals
\begin{eqnarray}\label{map1ints}
  I_1&=&\beta(x^2y+x^2z+xy^2+xz^2+y^2z+yz^2+2xyz)+\gamma(x^2+y^2+z^2)\nonumber
  \\
  & &{}+\epsilon(xy+xz+yz)+\xi(x+y+z) \\
  I_2&=&\beta^2(x+y)^2(y+z)^2+\beta(\epsilon-\gamma)[(x+y)^2(y+z)+(x+y)(y+z)^2]
  \nonumber \\
  & &{}+\gamma(\epsilon-2\gamma)[(x+y)^2+(y+z)^2]
  +[\beta\xi+(\epsilon-\gamma)(\epsilon-2\gamma)](x+y)(y+z)
  \nonumber \\
  & &
  {}+\xi(\epsilon-2\gamma)[(x+y)+(y+z)]
\end{eqnarray}
and
\begin{eqnarray}\label{map2}
  x'&=&y \nonumber \\
  y'&=&z \\
  z'&=&-x-
  \frac{2\beta yz+\epsilon(y+z)+\xi}
  {\alpha yz+\beta(y+z)+\gamma}\nonumber
\end{eqnarray}
with integrals
\begin{eqnarray}\label{map2ints}
  I_1&=&\alpha^2x^2y^2z^2+2\alpha \beta xyz(xy+xz+yz)+\alpha \epsilon xyz(x+y+z)\nonumber \\
  & &{}+ \beta^2(xy+xz+yz)^2+\beta \epsilon(x^2y+x^2z+xy^2+xz^2+y^2z+yz^2+4xyz)\nonumber \\
  & &{}+ (\alpha \xi-2\beta \gamma)xyz+\gamma \epsilon(x^2+y^2+z^2-xy-xz-yz)-\gamma^2 (x^2+y^2+z^2)
  \nonumber \\
  & &{}+ (\beta \xi+\epsilon^2)(xy+xz+yz)+\xi(\epsilon-\gamma)(x+y+z)\\
  I_2&=&(\beta+\epsilon)[\alpha xyz(x-y+z)+\gamma(x^2+y^2+z^2-xy-yz+xz)+\xi(x+z)]
  \nonumber \\
  & &{}+ [\beta^2(x+z)+\beta \epsilon(x+z+1)+\epsilon^2][xy+xz+yz-y^2].
\end{eqnarray}

We next consider the orientation-preserving case. Consider the
three-quadratic expression (\ref{threequad}) possessing the
symmetry $I(x,y,z)=I(y,z,-x)$ \footnote{This symmetry is due to
\cite{gm}.}. Following the procedure outlined above we obtain the
mapping
\begin{eqnarray}\label{3dvolpres}
    x'&=&y\\ \nonumber
    y'&=&z\\ \nonumber
    z' &=& x+\frac{A (y-z)}{B y z+C}
\end{eqnarray}
with integrals
\begin{eqnarray}\label{3dvolpres1}
    I_1&=&B^2 x^2 y^2 z^2-A B x y z (x-y+z) -A (A+C) (x y-x z+y z)
    \nonumber\\
    & &{}-C (A+C) (x^2+y^2+z^2)\\
    I_2&=&B^3 x^2 y^2 z^2 -C^3 (x y+x z+y z)+A B C (2 x z-2 x y^2 z-y)
    +A^2 (B-C) y^2\nonumber\\
    & &{}-B C^2 (x^2+y^2+z^2+x+y+z+x y z [x+y+z ]) \nonumber\\
    & &{}+2 A B^2 x y^2 z-B^2 C x y z (x y z+1)-A C^2 (x+y) (y+z)
\end{eqnarray}

We close this section with the following remark.

\vspace{3mm} {\bf Remark} The mapping (\ref{map1}) under the
coordinate transformation $X=x+y,Y=y+z$ can be reduced to a
two-dimensional area-preserving mapping, i.e.
\begin{equation}\label{map12d}
  L_1:\, \; \;X'=Y\, , \; \;Y'=-X-
  \frac{(\epsilon-\gamma)Y+\xi}
  {\beta Y+\gamma}
\end{equation}
with the second integral, $I_2$, becoming
\begin{eqnarray}\label{map12dints}
  I&=&\beta^2X^2Y^2+\beta(\epsilon-\gamma)(X^2Y+XY^2)+\gamma(\epsilon-2\gamma)(X^2+Y^2)
  \nonumber \\
  &
  &{}+[\beta\xi+(\epsilon-\gamma)(\epsilon-2\gamma)]XY+\xi(\epsilon-2\gamma)(X+Y).
\end{eqnarray}
Note that the first integral, $I_1$, does not reduce under this
transformation.

In fact, the mapping (\ref{map1}) is a member of a
recently-discovered hierarchy of integrable mappings given in
\cite{iat}, the three-dimensional asymmetric mapping\footnote{In
\cite{iat} this asymmetric mapping was shown to be obtained as a
composition of three involutions. We believe that this guarantees
the reversibility of this mapping. It seems, more generally, that
a mapping obtained from a composition of involutions is
reversible, see \cite{iat} for examples of such mappings.} being
\begin{eqnarray}\label{L2asym}
  x'&=&-x-\frac{\beta(y+z)^2+\epsilon(y+z)+\xi_0}{\beta(y+z)+\gamma_0}\nonumber
  \\
  y'&=&-y-\frac{\beta(x'+z)^2+\epsilon(x'+z)+\xi_1}{\beta(x'+z)+\gamma_1}\nonumber
  \\
  z'&=&-z-\frac{\beta(x'+y')^2+\epsilon(x'+y')+\xi_2}{\beta(x'+y')+\gamma_2}.
\end{eqnarray}

\section{Three-Dimensional Measure-Preserving Mappings}

In this section we apply the processes of reparametrization and
replacement to the three-dimensional volume-preserving integrable
mappings constructed above to construct measure-preserving
integrable mappings. These examples illustrate how integrable
three-dimensional families of mappings can be embedded in larger
(i.e. higher number of parameters) integrable three-dimensional
families of mappings via the processes of reparametrization and
replacement.

Consider the mapping (\ref{map2}) when $\alpha=0$ and $\epsilon =
\gamma$, i.e.
\begin{eqnarray}\label{map3}
  x'&=&y \nonumber \\
  y'&=&z \nonumber \\
  z'&=&-x-
  \frac{2\beta yz+\gamma(y+z)+\xi}
  {\beta (y+z)+\gamma}
\end{eqnarray}
which has integrals
\begin{eqnarray}\label{map3ints}
  I_1&=&\beta(xy+xz+yz)^2+\gamma(x+y)(x+z)(y+z)+\xi(xy+xz+yz)\\
  I_2&=&(\beta+\gamma)[\gamma(x^2+y^2+z^2-xy-yz+xz)+\xi(x+z)]
  \nonumber \\
  & &{}+ [\beta^2(x+z)+\beta \gamma(x+z+1)+\gamma^2][xy+xz+yz-y^2].
\end{eqnarray}
Notice that the parameters $\beta, \gamma$ and $\xi$ now appear
linearly in the integral $I_1$. Reparametrizing the parameters,
i.e. $\beta \rightarrow \beta_0 + \beta_1 K, \gamma \rightarrow
\gamma_0 +\gamma_1 K,\xi \rightarrow \xi_0 + \xi_1 K$ and the
integral $I_1 \rightarrow \bar{I_1}=I_1 + \mu_0 + \mu_1 K$, we
obtain the mapping
\begin{eqnarray}\label{map3a}
  x'&=&y \nonumber  \\
  y'&=&z \nonumber  \\
  z'&=&-x-
  \frac{2(\beta_0 + \beta_1 K) yz+(\gamma_0 +\gamma_1 K)(y+z)+\xi_0 + \xi_1 K}
  {(\beta_0 + \beta_1 K)(y+z)+(\gamma_0 + \gamma_1 K)}.
\end{eqnarray}
Using $\bar{I}_1(x,y,z)=0$ (as $\bar{I}_1(x,y,z)=0 \Rightarrow
\bar{I}_1(x',y',z')=0$) a new integral $K=k(x,y,z)$ can be
defined. Define the map $L_K$ to be the map (\ref{map3a}) with
replacement $K=k(x,y,z)$. The map $L_K$ has two integrals
$k(x,y,z)$ and $\bar{I}_2=I_2(x,y,z)|_{K=k(x,y,z)}$, i.e.
\begin{eqnarray}\label{map3aints}
  k&=&-\frac{\beta_0(xy+xz+yz)^2+\gamma_0(x+y)(x+z)(y+z)+\xi_0(xy+xz+yz)+\mu_0}
  {\beta_1(xy+xz+yz)^2+\gamma_1(x+y)(x+z)(y+z)+\xi_1(xy+xz+yz)+\mu_1}\nonumber\\
  & &\\
  \bar{I}_2&=&\{[\beta_0 +\gamma_0 + (\beta_1+\gamma_1) K][(\gamma_0+\gamma_1K)(x^2+y^2+z^2-xy-yz+xz)
  \nonumber \\
  & &{}+(\xi_0+\xi_1K)(x+z)]
  + [(\beta_0 + \beta_1 K)^2(x+z)\nonumber  \\
  & &{}+(\beta_0 + \beta_1 K) (\gamma_0 + \gamma_1 K)(x+z+1)
  \nonumber  \\
  & &{}+ (\gamma_0 + \gamma_1 K)^2][xy+xz+yz-y^2]\}|_{K=k(x,y,z)}.
\end{eqnarray}
The map $L_K$ is also measure preserving with
\begin{equation}
  m(x,y,z) = \left[ \frac{\partial \bar{I}_1 }{  \partial K}
  \right]^{-1}.
\end{equation}

Consider the mapping (\ref{map2}) when $\alpha=0$ and $\epsilon =
\gamma=\beta$, i.e.
\begin{eqnarray}\label{map6}
  x'&=&y \nonumber \\
  y'&=&z \nonumber \\
  z'&=&-x-
  \frac{\beta (2yz+y+z)+\xi}
  {\beta (y+z+1)}
\end{eqnarray}
which has integrals
\begin{eqnarray}\label{map6ints}
  I_1&=&\beta[(xy+xz+yz)^2+(x+y)(x+z)(y+z)]+\xi(xy+xz+yz)\\
  I_2&=&\beta(x+z)(x+z+xy+xz+yz-y^2)+\xi(x+z).
\end{eqnarray}
Notice that the parameters $\beta$ and $\xi$ now appear linearly
in both integrals. Reparametrizing the parameters and the
integrals, i.e. $\beta \rightarrow \beta_0 + \beta_1 K_1 + \beta_2
K_2$, $\xi \rightarrow \xi_0 + \xi_1 K_1 + \xi_2 K_2$,  $I_1
\rightarrow \bar{I}_1=I_1 + \mu_0 + \mu_1 K_1 + \mu_2 K_2$  and
$I_2 \rightarrow \bar{I}_2 = I_2 + \nu_0 + \nu_1 K_1 + \nu_2 K_2
$, we obtain the mapping
\begin{eqnarray}\label{map5}
  x'&=&y \nonumber \\
  y'&=&z \nonumber \\
  z'&=&-x-
  \frac{(\beta_0 + \beta_1 K_1 + \beta_2 K_2)(2yz+y+z)
  +\xi_0 + \xi_1 K_1 + \xi_2 K_2}
  {(\beta_0 + \beta_1 K_1 + \beta_2 K_2)(y+z+1)},
\end{eqnarray}
with integrals $\bar{I}_1(x,y,z)$ and $\bar{I}_2(x,y,z)$. Setting
$\bar{I}_1(x,y,z)=0$ and $\bar{I}_2(x,y,z)=0$ it is possible to
solve for $K_1=k_1(x,y,z)$ and $K_2=k_2(x,y,z)$ as $\bar{I}_1$ and
$\bar{I}_2$ are linear in $K_1$ and $K_2$. Define the map
$L_{K_1K_2}$ to be the map (\ref{map5}) with replacements
$K_1=k_1(x,y,z)$ and $K_2=k_2(x,y,z)$. The map $L_{K_1K_2}$ has
the integrals $K_1=k_1(x,y,z)$ and $K_2=k_2(x,y,z)$. The map
$L_{K_1K_2}$ is also measure preserving with
\begin{equation}
  m(x,y,z) = \left| \begin{array}{cc}
    \frac{\partial \bar{I}_1 }{  \partial K_1} & \frac{\partial \bar{I}_1 }{  \partial K_2} \\
    \frac{\partial \bar{I}_2 }{  \partial K_1} & \frac{\partial \bar{I}_2 }{ \partial K_2}  \
  \end{array} \right|^{-1}.
\end{equation}
The integrals to the above maps can be shown to be functionally
independent and in involution with respect to the following
Poisson structure \cite{bhq}
\begin{equation}\label{poisson}
  m(x,y,z)\left(\begin{array}{ccc}
    0 & \frac{\partial I }{ \partial z} & -\frac{\partial I }{ \partial y} \\
    -\frac{\partial I }{ \partial z} & 0 & \frac{\partial I }{ \partial x} \\
    \frac{\partial I }{ \partial y} & -\frac{\partial I }{ \partial x} & 0 \
  \end{array}\right),
\end{equation}
where $I$ is either one of integrals and $m(x,y,z)$ is the
measure.

Finally, we consider the mapping (\ref{map1}). As noted in the
remark at the end of Section 2 we can use a coordinate
transformation, i.e. $X=x+y$ and $Y=y+z$, to reduce the mapping to
a two-dimensional mapping. Importantly, however, the
three-quadratic integral, $I_1$, does not reduce under this
coordinate transformation and as a result if we use the processes
of reparametrisation and replacement on this integral then the
resulting mapping, $L_r$, is not reducible to a two dimensional
mapping, although for every fixed $K$ it is. The remark at the
end of Section 2 also shows that the reduced mapping has a
biquadratic integral and thus can be explicitly
integrated\footnote{This also is true for its asymmetric form.},
see \cite{aijr1}. This result can be used to integrate the
mapping $L_r$ also {\em but} this time curve-wise
(leaf-wise).\footnote{We believe that the maps considered in Case
2 of \cite[Section 3]{iat} can be integrated in this way also.}

\section*{Acknowledgements}

Part of this research was undertaken while in receipt of a La
Trobe University Postgraduate Research Scholarship.

\section*{Appendix }

Our work on multidimensional integrable mappings (particularly
\cite{iat}) has lead us to the following observation about the
degree of the variables that occur in the integrals :
\newline

{\em A 2n-dimensional volume-preserving integrable mapping with
variables $v_0,\dots,v_{2n-1}$, which has one $n$-quadratic
integral, possesses an additional $(n-1)$ integrals of the
following form}
\begin{equation}\label{intcon1}
  \begin{array}{c}
    I_2=\sum A_{\alpha_{21}\dots\alpha_{2,2n}} v^{\alpha_{21}}_0
  \dots v^{\alpha_{2,2n}}_{2n-1} \, , \; \;
  (\alpha_{21},\dots,\alpha_{2,2n}=0,\dots,4)\\
    \vdots \\
    I_n=\sum A_{\alpha_{n,1}\dots\alpha_{n,2n}} v^{\alpha_{n,1}}_0
  \dots v^{\alpha_{n,2n}}_{2n-1} \, , \; \;
  (\alpha_{n,1},\dots,\alpha_{n,2n}=0,\dots,2n).\
  \end{array}
\end{equation}

{\em While a (2n+1)-dimensional volume-preserving integrable
mapping with variables $v_0, \dots, v_{2n}$, which has one
$n$-quadratic integral, possesses an additional n integrals of the
following form}
\begin{equation}\label{intcon2}
  \begin{array}{c}
    I_2=\sum A_{\alpha_{21}\dots\alpha_{2,2n+1}} v^{\alpha_{21}}_0
  \dots v^{\alpha_{2,2n+1}}_{2n} \, , \; \;
  (\alpha_{21},\dots,\alpha_{2,2n+1}=0,\dots,4)\\
    \vdots \\
    I_{n+1}=\sum A_{\alpha_{n+1,1}\dots\alpha_{n+1,2n+1}} v^{\alpha_{n+1,1}}_0
  \dots v^{\alpha_{n+1,2n+1}}_{2n} \, , \; \;
  (\alpha_{n+1,1},\dots,\alpha_{n+1,2n+1}=0,\dots,2n+2).\
  \end{array}
\end{equation}
In the case we have considered in this paper, the power of the
first and last variables ranges from 0 to 2.

\end{document}